# Induced Ferromagnetism in bilayer Hexagonal Boron Nitride (h-BN) on vacancy defects at B and N sites

B. Chettri,[1, 2] P. K. Patra,[2] Tuan V. Vu,[3, 4, *] Lalrinkima,[5, 1] Abu Yaya,[6]
Kingsley O. Obodo,[7] Ngoc Thanh Thuy Tran,[8] A. Laref,[9] and D. P. Rai[1]

[1]*Physical Sciences Research Center (PSRC), Department of Physics, Pachhunga University College Aizawl 796001, India*
[2]*Department of Physics, North Eastern Hill University, Shillong , India*
[3]*Division of Computational Physics, Institute for Computational Science,*
*Ton Duc Thang University, Ho Chi Minh City, Vietnam*
[4]*Faculty of Electrical and Electronics Engineering, Ton Duc Thang University, Ho Chi Minh City, Vietnam*
[5]*Department of Physics, Mizoram University Aizawl 796009, India*
[6]*Department of Materials Science and Engineering, CBAS, University of Ghana, Ghana*
[7]*HySA Infrastructure Centre of Competence, Faculty of Engineering,*
*North-West University, South Africa (NWU), 2531, South Africa*
[8]*Hierarchical Green-Energy Materials (Hi-GEM) Research Center,*
*National Cheng Kung University, Tainan, 701, Taiwan*
[9]*Department of Physics and Astronomy, College of Science, King Saud University, Riyadh, 11451, Saudi Arabia*
(Dated: March 14, 2020)

We investigated the electronic and optical properties of bilayer AA' stacked Boron and Nitrogen vacancies in hexagonal Boron Nitride (h-BN) using density functional theory (DFT). The density of states (DOS) and electronic band structure showed that Boron vacancy in bilayer h-BN results in a magnetic and conducting ground state. The band gap energy ranges from 4.56 eV for the pristine BN bilayer to 0.12 eV for a single Nitrogen vacancy in the bilayer. Considering the presence of 1,3,4-Boron vacancy, half metallic character is observed. However, the 2-boron vacancy configuration resulted in metallic character. The bilayers with 1,2,3,4-Nitrogen vacancy has a band gap of 0.39, 0.33, 0.28 and 0.12eV respectively, which is significantly less than the pristine band gap. Also B and N vacancy induces ferromagnetism in the *h*-BN bilayer. The maximum total magnetic moment for the Boron vacant system is 6.583$\mu_B$ in case of 4-Boron vacancy configuration. In case of Nitrogen vacancy system it is 3.926$\mu_B$ for 4-Nitrogen vacancy configuration. The optical response of the system is presented in terms of the absorption coefficient, refractive index and dielectric constant for pristine as well as the deective configurations. Negative value of dielectric constant for Boron vacant system in the energy range 0.9-1.4 eV and for Nitrogen vacant system in the energy range 0.5-0.8 eV opens an opportunity for it to be utilized for negative index optical materials. The current study shows that B and N vacancies in bilayer h-BN could have potential applications in nano-structure based electronics, optoelectronics and spintronic devices.

## I. INTRODUCTION

The exfoliation of 2D materials and realization of its novel properties are opening a tremendous opportunities for advancement of future technology. Discovery of Graphene[1] and its importance for the advnacement of data storage and data communication technology has given a new dimension in the field of research in 2D nanomaterials. 2D graphene like materials such as MoS₂, ZnO, ZnS, CdO, etc., [2–8] are promising because of their wide band gap, nanoscale size compatibility, low weight and remarkable physical and chemical properties. Rai et al.[5] reported the effects of vacancy defect at S sites of 2D monolayer MoS₂ on the electronic and optical properties by using the computational package VNL-ATK based on the Density Functional Theory (DFT). One of such material is Boron Nitride (BN) which crystallizes in three phases: hexagonal Boron Nitride (h-BN), cubic Boron Nitride, and wurtzite Boron Nitride having a large band gaps in the range of 3.5-6.0 eV[9, 10]. The flexibility in modulating its electronic and optical properties by applying external electric field, strain, doping and creating defects as well as by chemical functionalisation is making this material a promising contender to revolutionize the spintronics, electronics and optoelectronics devices like LEDs, LASERs, Detectors, etc[11–13] and in many other fields[14]. The unique chemical, thermal , electrical, mechanical and structual properties of BN has diverse applications. Some of the important applications are dielectric subtrates as it greatly reduces the amplitude of the disorder-induced carrier density inhomogeneities [15, 16] or ultrathin seper-ation layers in graphene electronic devices[17]. BN nanosheets can be used as a wrapping shield for protecting of metals from oxidation and corrosion because of its great chemical and thermal stability[18]. Heterostructre bilayer of BN with graphene, silicene, germanene are satisfactory treated by using van der Waals (vDW) interaction to obtained the accuracy in electronic properties. This is because of its surfacial phenomena, large band gap, high thermal conductivity and can be fabricated as a gate dielectrics, electrodes, etc[19]. Monolayer and bilayer Boron Nitride are prepared by various experimental techniques such as the chemical vapour deposition (CVD), exfoliation techniques like: mechanical exfoliation and controllable electron beam irradiation[10, 20–24]. Sun et al.[25], successfully exfoliated 2-20 layers of Boron Nitride from bulk wurtzite BN powder in solution of thionyl chloride without using any dispersive agent. During the lab synthesis of Boron Nitride the defects like the vacancy defects, impurities[26–30], topological defects- specifically Stone Wales defects are observed. The presence of defects or impurities are due to the non-equillibrium microscopic growth via certain external perturbations, such as during rapid quench from high temperature or under irradiation[20], or by electron knock on processes[31], low energy ion bombardment[32] which affects the electronic properties as well. 2D materials obtained by stacking sequence generally formed a weak van der Waals bond, readily changes its electronic, thermal and mechanical properties which brings scientific inquisitiveness in studying and designing of multilayer materials for nanodevice applications[33]. Wang et al.[34] reviewed the structural and chemical stability of the promising 2D materials like h-BN, graphene and TMDs. In order to enhance the stability they have reported their possible work on surface modification. Constantine et al.[35] employed local second-order Moller-Plesser Perturbation Theory(LMP2) calculations for the solid state, as implemented in CRYSCOR and showed that the most stable stacking order is found to be AA' which relies on the interlayer distance and order of stacking incorporating London dispersion and electrostatic interactions. Zhang et al., using a dual-gate bilayer graphene field-effect transistor(FET) and infrared microspectroscopy demonstrated a gate-controlled, tunable band gap of bilayer graphene upto 250 meV[36]. Balu et al.[37] studied the electronic properties by applying the transverse effect of electric field on Graphene/BN and BN/BN bilayers using the full-potential linearized augmented plane wave (FLAPW) method within the framework of local density approximation (LDA-DFT). They have reported the increase in the band gap for Graphene/BN bilayer and reverse phenomena for BN/BN bilayer under applied transverse electric field. The implementation of vDW force in multilayer hexagonanal 2D materials (h-BN) is highly important for holding the layers at their fixed positions[38]. The study of the twisted bilayer graphene and twisted bilayer Boron Nitride with vDW interactions has shown dramatic changes in optical and excitonic properties[39]. The response of the tunneling magnetoresistance of the bilayer h-BN based magneto-tunnel junction (MTJ) had been examined under external uniaxial strain and it has shown a decent result to become a novel component for MTJs[40]. Fartab et al. [41] studied the effect of the vacancy on monolayer h-BN sheet with and without Lithium doping using Plane Wave pseudopotential based DFT as implemented in Quantum Expresso package. It has been reported that the B and N vacancies, induces spontaneous magnetisation with a total magnetic moment of $2.81\mu_B$ and $0.98\mu_B$ for single B and single N vacancy defects, respectively. While substitution of Boron or Nitrogen by Li-atom gives the total magnetic moment of $2.0\mu_B$. Huang et al., has also reported the similar fact about induced spin moment on the effect of defect and impurity on the monolayer h-BN from fist principles method. They have also reported the senstivity of bond length with respect to charge states [28]. A DFT study on the hydrogenated Boron Nitride nanosheets and nanoribbons with different widths have been studied and reported the presence of ferromagnetic behaviour in the ziz-zag BNNR, whereas the armchair is a non-magnetic semiconductor[42]. Since decades, rigorous researches are in progress from both theory and experiment, in order to explore the most stable multilayer arrangement of the Boron Nitride by varying the nature of stacking in different alternate patterns[43, 44]. Recently, presented the electronic properties of BN thin films by using DFT under $sp2$ and $sp3$ hybridization with possible configurations of AA, AA', AB, AB' stackings[46]. Fujimoto et al. reported the variation of band gaps in bilayer h-BN depends on the nature of layer stacking, interlayer distance between them and strain(bi-axial/uniaxial)[47]. It is found that h-BN shows lattice compatibility with other 2D materials like Graphene, $MoS2$ etc., and heterostrutures have been studied for practical applications[48, 49]. The study of silicene and h-BN heterostructure bilayer shows remarkable thermoelectric properties[50]. In this paper, we have presented the electronic properties of the defects in hexagonal Boron Nitride bilayer with AA' (Boron on top of Boron and Nitrogen on top of Boron). This is the most stable configuration among all the combinations as reported realier[43–45]. The effect of defects on bilayer h-BN has not been reported earlier to the best of our knowledge. Therefore, we intend to carry forward this work and the results can be useful for experimental references as well as for the development of technological devices.

* Corresponding Author:vuvantuan@tdtu.edu.vn (Tuan V. Vu)

## II. COMPUTATIONAL DETAIL

The electronic and optical calculations were performed by using the Density Functional Theory (DFT)[51] within Linear Combination of Atomic Orbital (LCAO) method as implemented in the Quantumwise VNL-ATK[52]. The bilayer BN system is created by placing the single layer of BN one above the other in AA and then rotated it to make it AA' with the interlayer spacing taken as 3.34Å. The vacuum of 15 Å is applied to break the transversal symmetry along the z-axis which avoid the interaction of bilayers. All the electron-electron interactions were considered using spin generalized gradient approximation with Perdew-Burke-Ernzerhof(PBE)[53]. We have also considered the van der Waals (vdW) interactions[54] between the two adjacent layers within the Grimme DFT-D2 correction parameters[55] for the description of accurate electronic properties. The electron-ion interaction is treated by using the projected augmented plane wave (PAW) method. A model consisting of Boron Nitride bilayer were formed with 32 atoms in each layer. The geometry optimization was carried out until maximum force component on each individual constituent atom reached less than 0.0005eV/Å. To analyse the modulation of electronic and optical properties we have used atomic site defects or add impurities method [2, 5]. The atomic sites vacancies were created by random removal of the Boron and nitrogen atom from their respective sites. The concentration of defect is presented as 0.03125% (x=1/32) which signifies 1 atom out of 32 Boron/Nitrogen atom was removed and similarly removal of 2,3 and 4 vacancy defects are given as x=0.0625%, 0.09375% and 0.125% respectively. The phonon band structure was calculated to examine the thermodynamical stability. For all electronic calculations, we have integrated the first Brillouin zone by taking the k-mesh 16×16×1 within Monkhorst package[56].

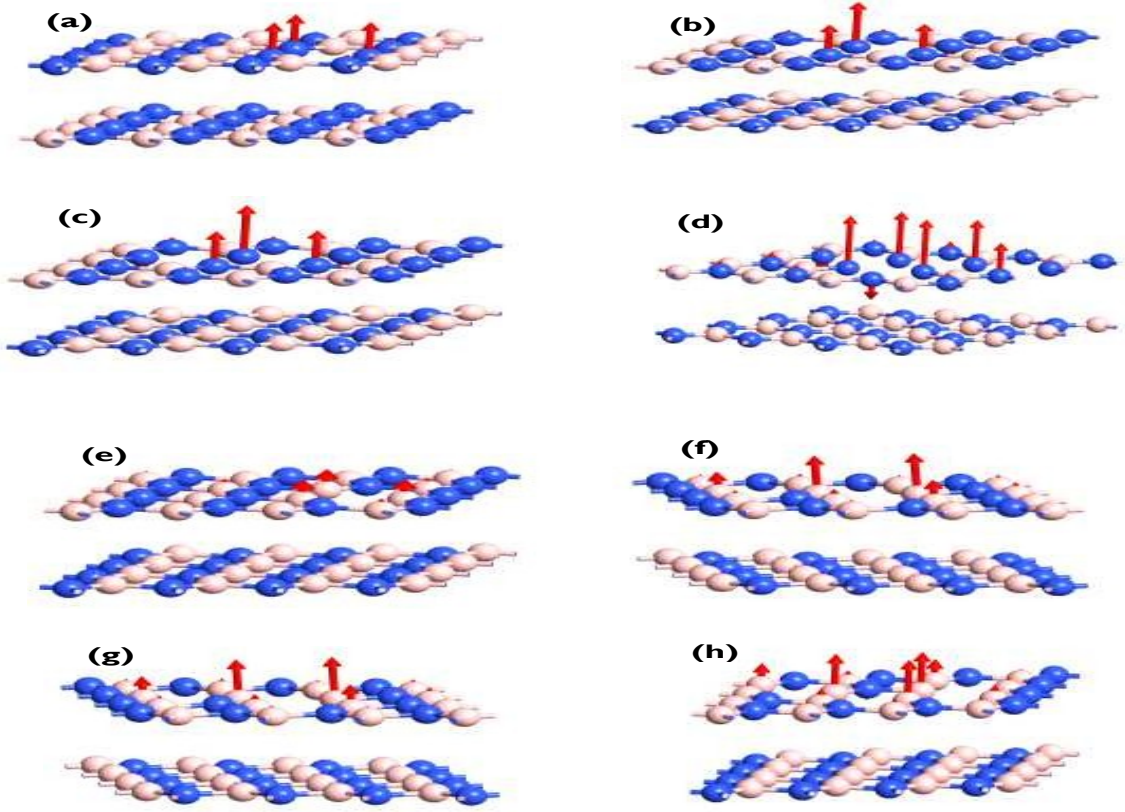

FIG. 1: . Spin magnetic moment of vacant system labelled (a)-(d) for Boron vacancy and (e)-(h) for Nitrogen vacant system respectively with 0.03125% ,0.0625 , 0.09375% , 0.125% vacancy. The red arrow denotes the magnetic moment of the B/N atoms at the vacancy sites.

## III.   RESULTS AND DISCUSSION

Fig. 1 shows the optimized bilayer structures and the respective bond lengths between N-N, B-B and B-N or N-B are 2.50Å, 2.50Å and 1.45Å in the pristine sysetem, consistent with the previous studies[57]. The optimized interlayer distance between the two BN sheets is found to be ∼3.34Å which is in good agreement with the previous findings[35, 58–63]. We have created a Boron and Nitrogen vacancy defects on the top layer [see Fig.1(a-h)]. The increase in number of defects at B sites increases the N-N, B-B, N-B bond lengths by ∼0.01Å. However, in the case of N site defects the N-N and B-B bond length remains unchanged but B-N or N-B decreases by ∼0.01Å. Further, we have calculated the formation energies ($E_f$) of each system from Eq.(1)[64] to test the structural stability. The calculated formation energies are presented in Fig.2(a,b).

$$E_f = \frac{E_{tot.} - (n_1 E_B + n_2 E_N)}{(n_1 + n_2)} \tag{1}$$

where $E_B$, $E_N$ , $n_1$, $n_2$ are the total ground state energies of isolated Boron, Nitrogen, total number of Boron and Nitrogen atoms, respectively. For the final confirmation of thermodynamical stability the phonon dispersion bands are calculated for all the systems as presented in Fig. (3, 4). The large number of optical modes are obtained above 25 meV and acoustic modes are below 25 meV. It is clrealy seen that the frequencies lie above zero in the positive quadrant. This proves that all the systems under our investigations are thermodynamically stable and can be practically synthesize in lab. We have shown the phonon dispersion only upto 0 to 4 vacancies this is becasue the phonon above 4 atomic(B/N) vacancy defects are highly unstable with range of negative frequency modes.

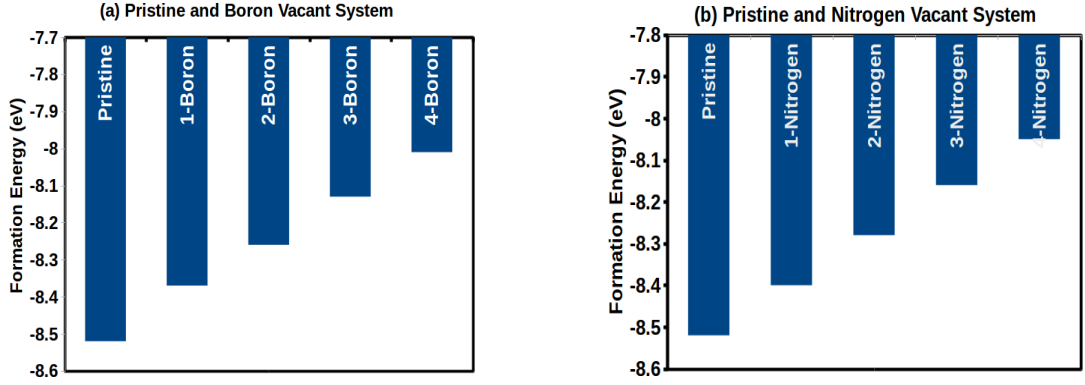

FIG. 2: Calculated formation energies of all the B and N vacancy defects: (a)-Pristine and B vacancy systems. (b)-Pristine and N vacancy systems. (Number of B and N vacancy defects are denoted by 1-Boron, 1-Nitrogen and so on respectively)

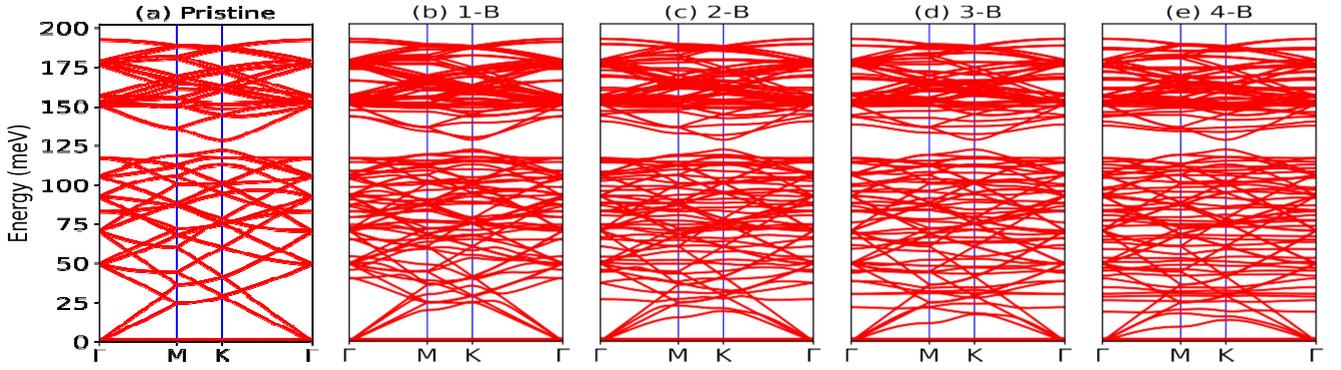

FIG. 3: Phonon dispersion band structure with number of Boron vacancies (a) Pristine(x=0.00%), (b) x=1.0 (0.03125%) (c) x=2.0 (0.0625%) (d) x=3.0(0.09375%) and (e) x=4.0 (0.125%)

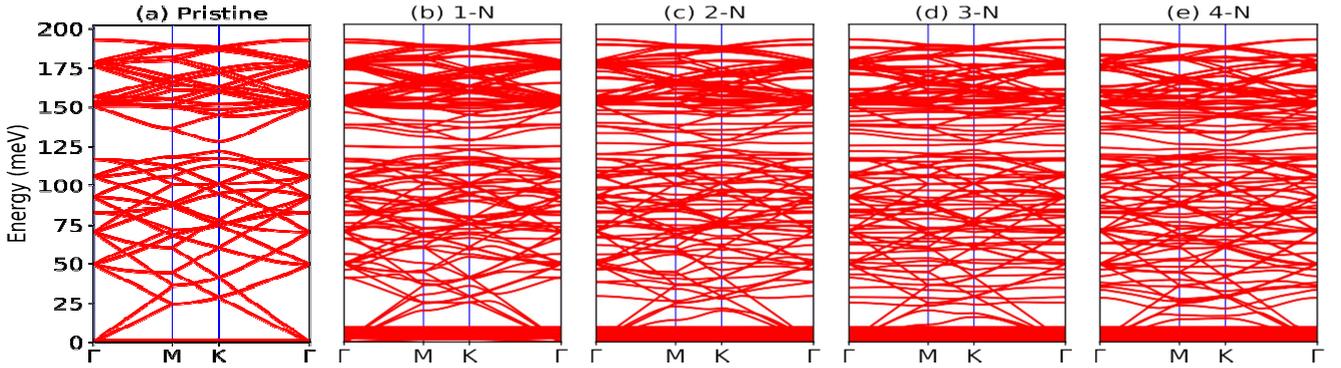

FIG. 4: Phonon dispersion band structure with number of Nitrogen vacancies (a) Pristine(x=0.00%), (b)x=1.0 (0.03125%) (c) x=2.0(0.0625%) (d) x=3.0(0.09375%) and (e) x=4.0 (0.125%)

## A. Electronic properties

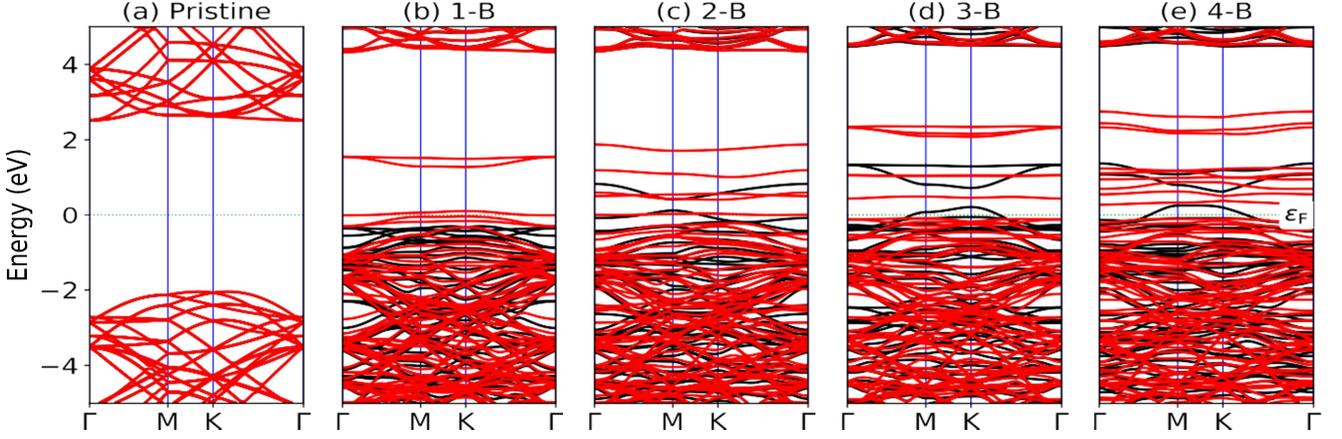

FIG. 5: . Electronic band structure with number of Boron vacancies (a) Pristine(x=0.00%), (b) x=1.0 (0.03125%) (c) x=2.0(0.0625%) (d) x=3.0(0.09375%) and (e) x=4.0 (0.125%). Here black and red coloured lines denotes spin-up and spin-down band lines respectively.

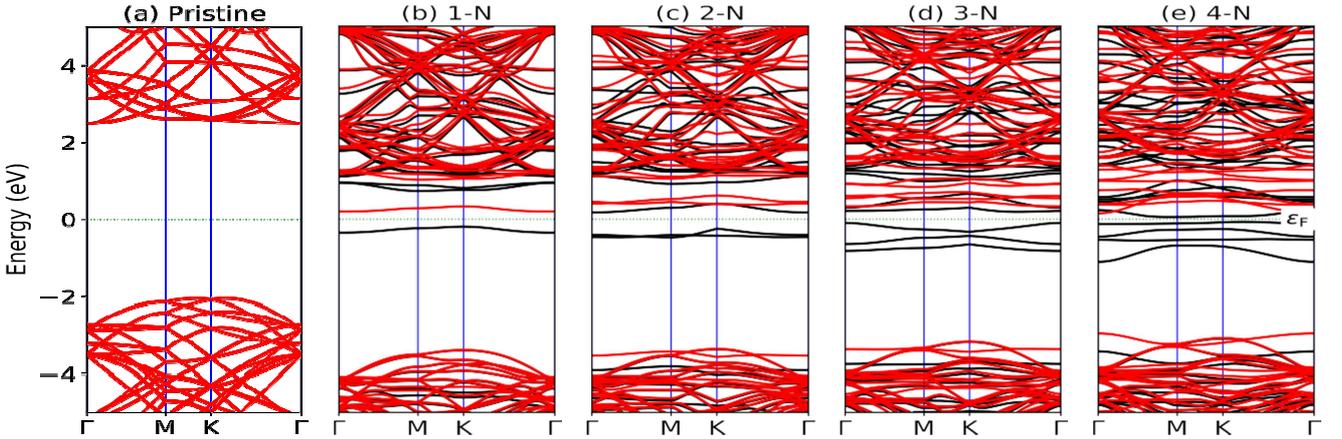

FIG. 6: . Electronic band structure with number of Nitrogen vacancies (a) Pristine(x=0.00%), (b) x=1.0 (0.03125%)(c) x=2.0(0.0625%) (d) x=3.0(0.09375%) and (e) x=4.0 (0.125%). Here black and red coloured lines denotes spin-up and spin-down band lines respectively.

Fig. 5(a), illustrates the electronic band structure of the pristine bilayer h-BN system which exhibit a perfect symmetry between the spin-up and the spin down bands indicating the antiferromagnetic behaviour. The calculated band gap was found to be ∼4.56eV for the pristine h-BN bilayer system($x = 0.00$) which can also be cross verified from the density of states (DOS) as well [see supporting figures Fig-s.14(a)]. Our results of electronic band structure and the calculated band gap is in good agreement with other previous theoretical results[9, 35, 42, 47, 58–62]. It is obvious that the effect of B/N atom vacancies changes the exisisting electronic property of the system. The variation of electroic properties may give new opportunities for the advancement of new technological applications. Therefore we proceed our calculation by creating a vacancy site defect at B/N sites [see Fig.1]. We have observed the reduced band gap with anti-symetric DOS on vacacy defects at B and N sites [see supporting figures Fig-s.14(a,b)]. The anti-symmetric DOS indicates the presence of magnetic behaviour. The semiconducting system with finite magnetic behaviour is crucial in present scenario for the development of spintronic devices. Thus, our systems look promising in this regard as a magnetic semiconductors. A single defect at B site break the bond between B and N which creates single unbonded N atom denoted as N-1UBD (UBD refers to unbonded) [cf Fig.7(a)]. In two B atom vacacny defects there will be singly unbonded (N-1UBD) and double unbonded (N-2UBD) N atoms [cf. Fig.7(b)]. In three B atom vacacny defects we can see single singly unbonded, two double unbonded N atoms [cf. Fig.7(c)]. For four B atom vacancy defects there

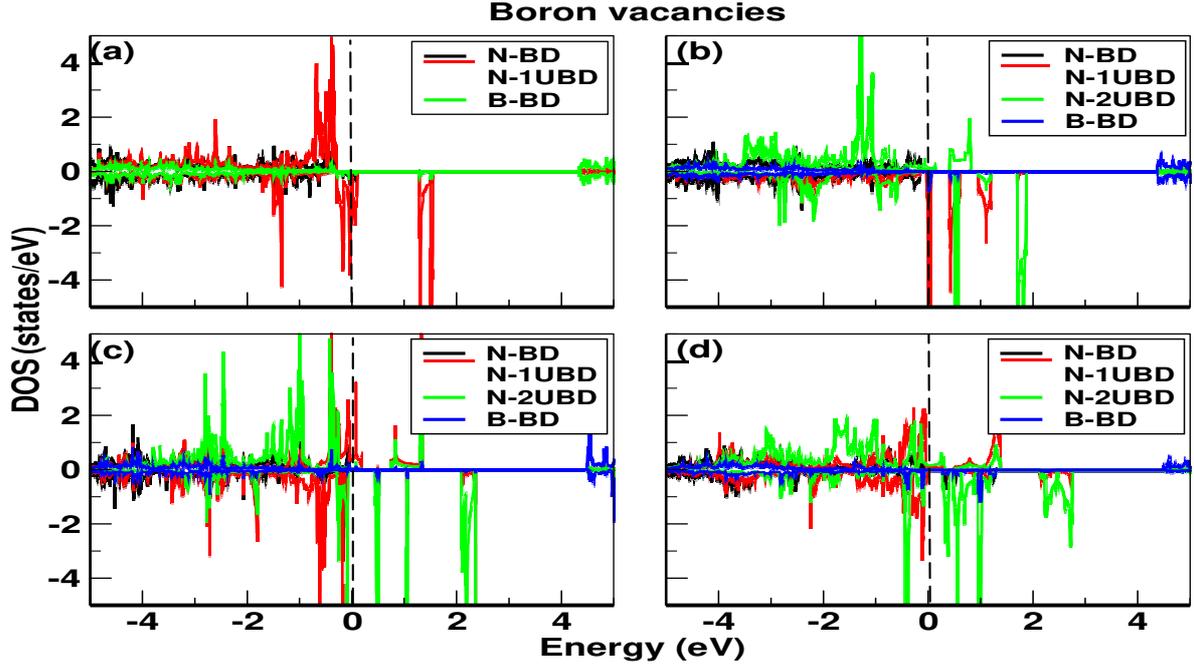

FIG. 7: Partial DOS of vacancy defects at B-sites (a) one vacancy (b) two vacancies (c) three vacancies (d) four vacancies. Fermi level set to zero. Here, BD- Bonded atoms, UBD-Unbonded atoms denotes the number of broken bonds of the Boron and Nitrogen atoms in vacant sites.

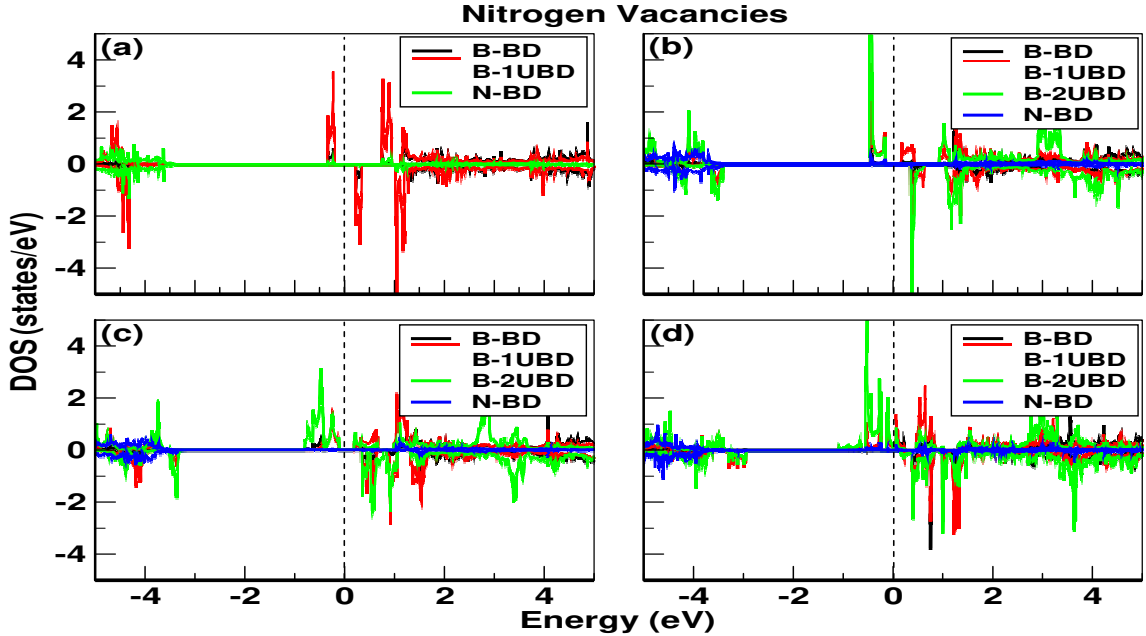

FIG. 8: . Partial DOS of vacancy defects at N-sites (a) one vacancy (b) two vacancies (c) three vacancies (d) four vacancies. Fermi level set to zero. Here, BD- Bonded atoms, UBD-Unbonded atoms denotes the number of broken bonds of the Boron and Nitrogen atoms in vacant sites.

are single and double unbonded N atoms as shown in Fig.7(d). Similarly, we can see the similar trend of unbonded bond length at B-site in the case of 1,2,3,4 N atom vacancy defects [cf Fig.8(a-d)].

The vacancy defects at B/N sites not only modify the electronic properties but also induces the ferromagnetic behaviour in bilayer BN. We have noticed two types of phase transition (i) nonmagnetic to magnetic (ii) Semiconductor to Metal to Half Metal. The co-existing of magnetic and semiconducting behaviour is interesting and can be of great importance in future data-transfer and data-storage technology. Single B-atom vacancy have an anti-symmetric DOS profile with the semiconducting band gap in both spin channels, thus resulting in a magnetic semiconductor. However, the spin down band gap is narrower than the gap at spin up channel due to dominant feature of the half filled N-2p orbital. A vacancy defect create unbonded bond length at N-site generating 3 free unpaired electrons on the neighbouring 3 N atoms and behave like a magnetic diople moment as shown by the arrow head in Fig.1(a). The increase in the B-atom vacancy results in the increased number of unpaired electrons at N-sites modifying the electronic properties of the system. In case of two B atom vacancy defects, single and double unbonded N atoms are formed. The system is a magnetic metal without diminished band gap at both the spin channels as the Fermi energy passes through the N-p states [Fig.1(b)]. The double unbonded N atom release more unpaired free electrons as compared to single unbonded N atom as a result the strength of moment at the former is more (the length of arrow head in N-2UBD is bigger than the arrow head in N-1UBD) [Fig.1(b)]. On the other hand, the three and four B vacancy defects create a metallic spin up and semiconducting spin down characteristics leading to large spin polarisation at the Fermi level, a finger print of a half-metal ferromagnetic(HMF). The calculated total and partial magnetic moments are presented in TableI. Our results of magnetic moments are compared with that of single layer BN with vacancy defects[41] due to the unavaiability of reported results of vacancy defects in BN bilayer [see TableI]. Similarly, for the N vacancy defects the bonds are broken at the B sites. As the electronic configuration of Boron is $1s^2 2s^2 2p_1$, therefore the single vacancy at N-site creates a spin state with a partial magnetic moment of 0.260 $\mu_B$ at each unbonded B-atom. The calculated total magnetic moment~ 0.903 $\mu_B$ and the partial magnetic moment~ 0.260 $\mu_B$ are in good agreement with the previous results[28, 41]. In case of the single Nitrogen vacancy two spin states i.e., 1 spin up in the valence band and 1 spin down in the conduction band are shifted near the fermi level[see Fig.6(b)]. In case of 2-N vacancy 4 single and 1 double unbonded B atoms are created at the B-site. These unpaired electrons of the unbonded Boron atoms introduced 3 spin up and 2 spin down states near the fermi level. As the double unbonded Boron atom has more number of unpaired electron as compared to single unbonded Boron atom, the strength of magnetic moment of the double unbonded B-atoms at the Nitrogen vacant site is greater in comparison with the single unbonded B-atom[cf Fig.1(f)]. Similarly for 3, 4-Nitrogen vacancies more number of single and double unbonded Boron atoms are created. Hence, increased number of the unpaired electrons on the Boron atoms at the vacant site. Thus, the number of unpaired electrons in the B-atom during the 1,2,3,4-Nitrogen vacancies leads to the tuning of the band gap of the Nitrogen vacant system[Fig.6]. Moreover, the calculated band gap for 1,2,3,4-Nitrogen vacancies are 0.39, 0.33, 0.28 and 0.12eV respectively exhibiting a semiconducting behaviour. After succesive increase in Boron and Nitrogen vacanies, it is observed that the total magnetic moment in both B and N system increases[see TableI]. It can be inferred from the Fig.[1],[Table I] that the total magnetic moment of the B and N vacant system are mainly due to the contributions of single and double unbonded $2p_1$ and $2p_3$ orbitals of the Boron and Nitrogen atoms[Fig.(7,8)].

TABLE I: Total and Partial Magnetic moments ($\mu_B$) of BN bilayer with vacancy defects at B and N sites. The unbonded bondlegths are represented by singly unbonded (*-1UBD) and doubly unbonded (*-2UBD), *=B/N

| Defects | Site | Total | N-1UBD/B-1UBD | N-2UBD/B-2UBD | Previous |
|---------|------|-------|---------------|---------------|----------|
| 1 | Boron | 2.953 | 0.803 | xxx | 2.81[41] |
| 2 | Boron | 3.144 | 0.877 | 1.400 | xxx |
|   |       |       | -0.182 |      |     |
| 3 | Boron | 3.745 | -0.436 | 1.548 | xxx |
| 4 | Boron | 6.583 | 0.893 | 1.553 | xxx |
|   |       |       | 0.031 | 1.544 |     |
|   |       |       | -0.458 | 1.430 |     |
|   |       |       | -0.274 | 1.290 |     |
| 1 | Nitrogen | 0.903 | 0.260 | xxx | 0.98[41] |
| 2 | Nitrogen | 1.973 | 0.436 | 0.994 | xxx |
|   |          |       | 0.436 |      |     |
|   |          |       | -0.041 |      |     |
| 3 | Nitrogen | 2.975 | 0.366 | 0.927 | xxx |
|   |          |       | 0.399 | 0.845 |     |
|   |          |       | 0.119 |      |     |
| 4 | Nitrogen | 3.926 | 0.365 | 0.902 | xxx |
|   |          |       | 0.366 | 0.846 |     |
|   |          |       | 0.141 | 0.846 |     |

Electron localisation function (ELF) is presented in the Fig.9 for all the system. We have taken the value between 0.7-0.9 which is good enough to find the regions of the localised electrons for all the system. Boron and Nitrogen atoms in the Boron Nitride are covalently bonded via $sp_2$ hybridisation. From the Fig. 9(a) of pristine bilayer system the localisation of the electron in between the BN bonds resembles the convalent type of bonding. Further analysis of the Boron and Nitrogen vacant systems is presented in Fig.9(b-i) at the vacant site. It is observed that the charges are equally localised in the bonded BN bonds whereas there is increase in localisation of the electron on the nearby (1,2-UBD)unbonded Nitrogen and Boron atoms in case of Boron and Nitrogen vacancies[cf Fig.9(b)-(e)]. It is evident that the localisation will greater on the double unbonded B and N atoms in comparision with the single unbonded atoms as observed in fig.[9-(c)-(d)-(e)] and [9-(g)-(h)-(i)] for 2,3,4-Boron/Nitrogen vacant systems. It is because of the greater number of free unpaired electrons on the double unbonded B and N atoms during Nitrogen and Boron vacancies.

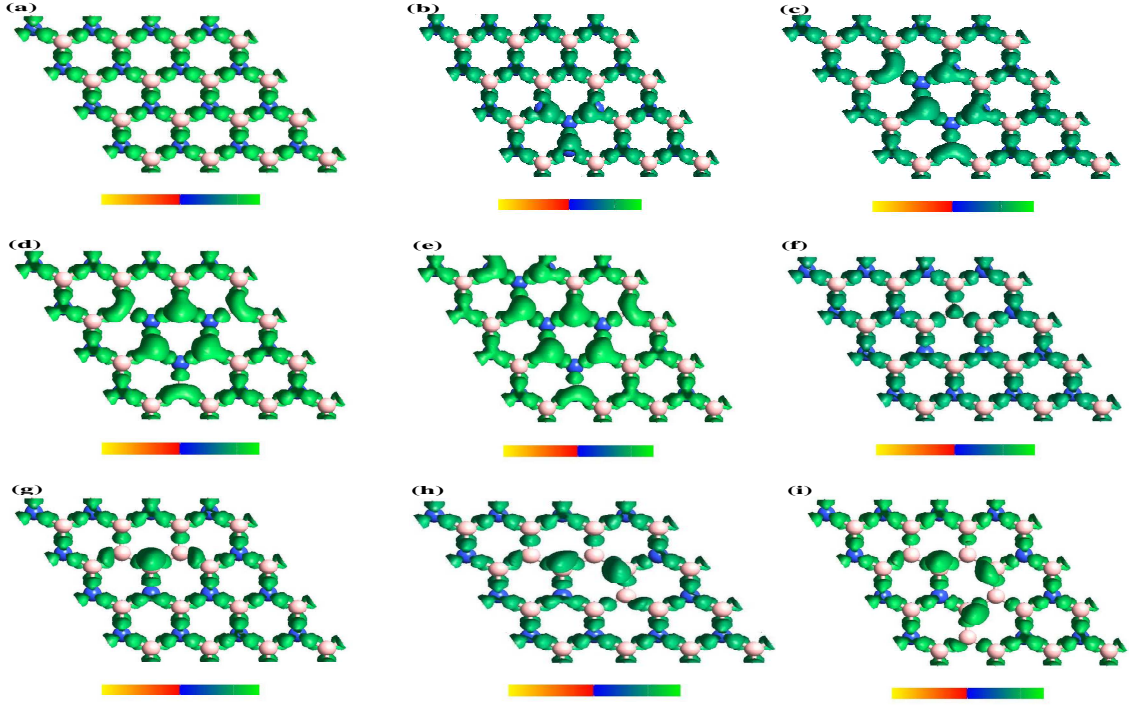

FIG. 9: Electron Localisation Function. It's range is from 0.0 - 1.00. Fig.(a) x = 0.00%, (b) 0.03125%, (c) 0.0625%, (d) 0.09375% and (e) 0.125% are for Pristine Bilayer, 1,2,3,4-Boron vacant system and Fig.(f) 0.03125%, (g) 0.0625%, (h) 0.09375%, (i) 0.125% are for 1,2,3,4-Nitrogen vacant system respectively.

## B. Optical Properties

Bilayer BN system without any defect shows a band gap of 4.56eV having wavelength equivalent as 275nm which falls in the Ultraviolet region of the Electromagnetic spectra. Out of the total electromagnetic radiation coming from the sunlight 10% of it is UV radiation. So materials having optical response in this range have been utilised for many practical purpose such as optical waveguide, radiation detectors etc. Thus, in this part of the paper the optical response has been analysed by calculating the properties like absorption coefficients, dielectric constants and reractive index as a function of energy of pristine x=0.00% and all the B and N vacancy system with defect percentage as x = 0.03125%, 0.0625%, 0.09375% and 0.125%. The response of a system to an electromagnetic radiation can be expressed in terms of dielectric function which is a complex quantity. The real part and imainary part of it can be expressed as,

$$\varrho = \varrho_1 + \iota\varrho_2 \qquad (2)$$

where $\varrho_1$- real and $\varrho_2$- imaginary part of the dielectric function respectively. The imaginary part of the dielectric function can be obtained by calculating the momentum matrix element between the occupied and unoccupied eigenstates given by Eg.(2)[66],

$$\varrho_2(\omega) = \frac{k^2 e^2}{\pi m^2 \omega^2} \Sigma_{mn'} \int_k d^3k \; |<\vec{k\,n}|\;\vec{p}\;|\vec{k\,n}'>|^2 [1 - f(\vec{k\,n})]\delta(\vec{E_{k\,n}} - \vec{E_{k\,n'}} - k\omega) \qquad (3)$$

where $\overrightarrow{p}$ - momentum operator, $\mid \overrightarrow{k}\,n >$ - eigenfunction of eigenvalue, $f(\overrightarrow{k}\,n)$- Fermi Distribution function. The real part of the dielectric function is obtained by the Kramers-Kronig transformation from its corresponding imaginary part as :

$$\varrho_2(\omega) = 1 + \frac{2}{\pi}\int_0^\infty \frac{\varrho_2(\omega')\omega'd\omega'}{\omega'^2 - \omega^2} \qquad (4)$$

The absorption coefficients $\alpha(\omega)$, refractive index $n(\omega)$ which are related to the dielectric function are given as follows;

$$\alpha(\omega) = \frac{2\omega\kappa(\omega)}{c} \qquad (5)$$

where $\kappa(\omega)$ is the extinxtion coefficient (imaginary part of complex refractive index).

$$n(\omega) = \sqrt{\frac{(\varrho_1 + \varrho_2)^2 + \varrho_1}{2}} \qquad (6)$$

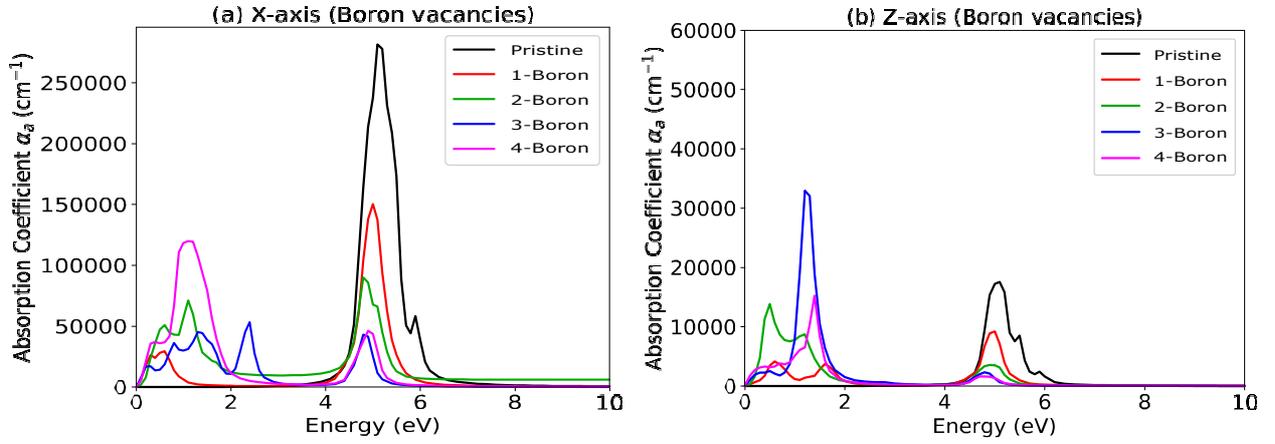

FIG. 10: Absorption coefficients with the labelling of the respective pristine(x=0.00%)-Fig(a) and Boron vacant bilayer system with x=0.03125%-Fig(b), 0.0625%-Fig(c), 0.09375%-Fig(d) and 0.125%-Fig(e) are presented in the above figures.

To give an overview of the optical properties of the pristine, B and N vacant systems, the calculated absorption coefficient, dielectric functions and refractive index with respect to photon energy are shown in Fig.10, 11, 15 for B=vacancies and 12, 13, 16 for N-vacancies with electric field parallel(along x-axis) and perpendicular(along z-axis) to the bilayer h-BN sheet. Also, the value of static dielectric constant, refractive index for pristine, 1,2,3,4-(Boron and Nitrogen) vacancies along both parallel and perpendicular polarizations are summarised in tableII. From the Fig.10(a) two absorption coefficient peak at two different energies are observed. Maximum absorption peak is at 5.1eV with a minimum peak at 5.9eV for the pristine(x=0.00%) bilayer Boron Nitride system. The calculated absorption coefficient for the pristine bilayer is $2.815 \times 10 \mathrm{s cm^{-1}}$. In Fig.10(b), for 1 Boron vacant system first absorption peak is observed at 0.75eV whereas a highest peak occurs at 4.90eV. With the increase in Boron vacancies a shift in the absorption peaks towards the lower energy are observed. In FIG.11, with the increase in the defect percentage, a gradual rise in the static dielectric constant at zero frequency/0.00eV can be observed. Fig.11-(a) shows the value of static dielectric constant for pristine(x=0.00%) bilayer along xx and zz directions as 1.113 and 1.001respectively. As we approach the higher energy the static dielctric constant along $\varrho_{xx}$ reaches zero for 4B vacancy at energy 1.0eV and it becomes negative due to the intraband transition in between 1.0-1.5eV thus enabling the occurence of plasmonic vibrations at this energy range[5, 67]. Whereas, for $\varrho_{zz}$ all values of the static dielectric constant remains positive corresponding to interband transitions with maximum value of 2.48 for 4B vacancy system[67, 68]. The imaginary part $\varrho_{2xx}$ and $\varrho_{2zz}$ of the dielectric function are presented in Fig.[11-(a-ii), (b-ii)] for Boron vacancies. A peak is observed at 5.1ev for $\varrho_{xx}$ pristine bilayer. For 2-B vacancies, one maximum and one minimum peaks are observed at 5.0eV and 0.2eV respectively. For 2-B vacancies 3 peaks are observed, maximum at 0.4eV and two minima at 1.0 and 4.8eV respectively. For 3-B vacancies 5 peaks are observed and the maximum is at 0.1eV. A maximum peak is observed at 0.25eV for 4-B vacancies with 2 minimas at 1.0eV and 4.6eV.

The calculated refractive index according to equation (5) for pristine and h-BN vacant bilayer system along x and z directions are presented in the supporting figures [see supporting figure Fig-(a,b).15]. Static refractive index for the pristine bilayer along x and z are found to be 1.063 and 1.001 respectively. For the pristine, maximum peak is observed at energy 4.6eV and the value of refractive index goes below 1 at 5.1eV showing transparent nature of the material at this energy range. The value of refractive index for 1 and 2 h-BN vacancy goes below 1 at 0.5eV, 1.0 respectively. Refractives index$\eta_{xx}$ goes below 1 twice, at 0.5eV and at 1.0eV for 4-h-BN Vacancy. In comparision with the pristine bilayer, in h-BN vacancy multiple peaks at different energies are observed.

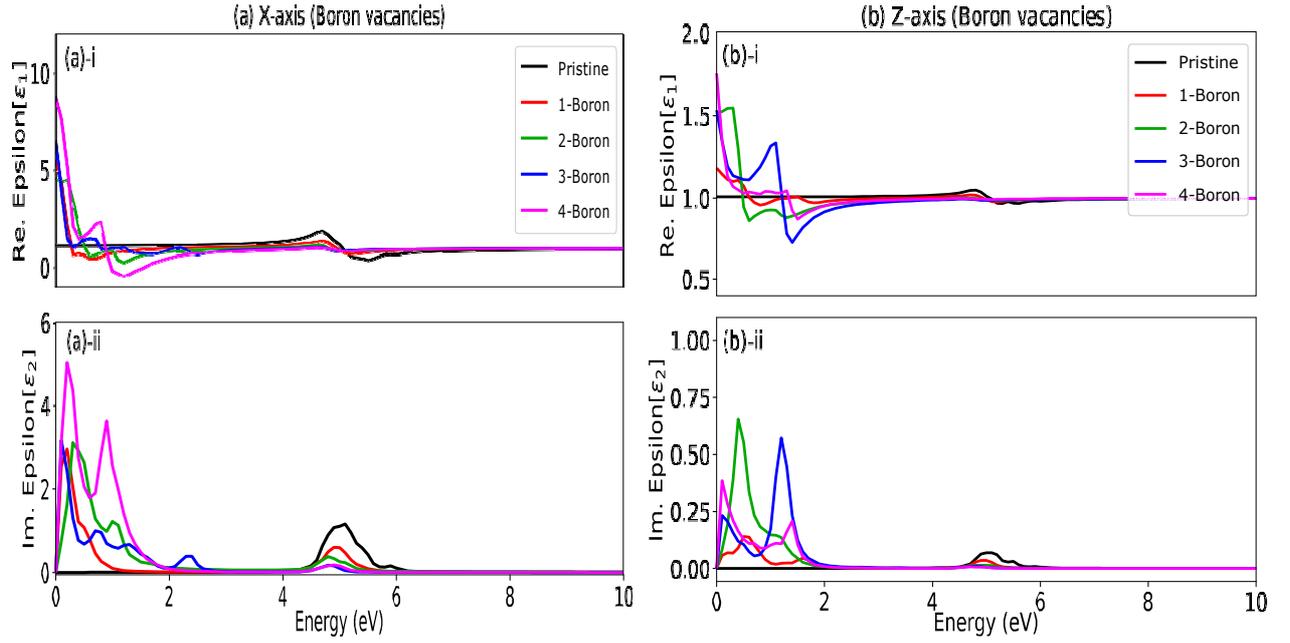

FIG. 11: The real and the imaginary part of the calculated dielectric function of the Pristine and Boron vacant system. Fig.(a)(b) is for in plane symmetry i.e., along xx. Fig.(c) is for out of plane symmetry i.e., along zz.

TABLE II: The calculated static dielectric constant, refractive index and the bandgap energy for the pristine and all other B and N vacant system.

| System | $\varrho_{1xx}$ | $\varrho_{1zz}$ | $n_{1xx}$ | $n_{1zz}$ | $E_g$ (eV) |
|---|---|---|---|---|---|
| Pristine | 1.113 | 1.001 | 1.063 | 1.001 | 4.56 |
| 1 Boron | 5.272 | 1.140 | 2.337 | 1.080 | 0.00 |
| 2 Boron | 4.566 | 1.544 | 2.131 | 1.226 | 0.00 |
| 3 Boron | 6.446 | 1.517 | 2.540 | 1.243 | 0.00 |
| 4 Boron | 8.695 | 1.718 | 2.926 | 1.322 | 0.00 |
| 1 Nitrogen | 1.212 | 1.011 | 1.098 | 1.002 | 0.39 |
| 2 Nitrogen | 1.215 | 1.077 | 1.093 | 1.028 | 0.33 |
| 3 Nitrogen | 2.239 | 1.173 | 1.494 | 1.081 | 0.28 |
| 4 Nitrogen | 3.786 | 1.599 | 1.946 | 1.262 | 0.12 |

The absorption coefficient of the N-vacancies along x and z are shown in Fig. 12.In absorption spectra12-(a), the maximum peak for 1,2,3,4-Nitrogen vacancies are observed at 5.2ev, 4.8eV, 4.1eV and 3.6eV respectively. The peaks are shifted towards the lower energies with increase in the number of N-vacancies. Similarly, the absorption spectra of N-vacancies along z axis is presented in Fig.12-(b). The peaks are observed at similar energy range which are shifted towards the lower energies with the increase in vacancies but the value of absorption coefficient along z axis is less in comparision with the absorption coefficient along x axis[12-(a),(b)].

The real($\varrho_1$) and the imaginary($\varrho_2$) dielectric function of Nitrogen vacant system for both polarization i.e., parallel(x) and perpendicular(z) to the plane of N-vacant h-BN bilayer is demonstrated in Fig.[13-(a,b)]. From the fig.[13-(a)-i] the values of static dielectric constant at 0.00eV for the 1,2,3,4-Nitrogen vacancies when the field is parallel(x) and perpendicular(z) are 1.212, 1.215, 2.239, 3.786 and 1.011, 1.077, 1.173, 1.599 respectively. The real part of the dielectric function becomes negative in case of 4-nitrogen vacancies in the energy range 0.677-0.897eV. With the increase in the vacancies, the peaks in imaginary part [13-(a)-ii, (b)-ii] of the dielectric function for 1,2,3,4-Nitrogen vacancies are shifted towards the lower energies(frequencies). The imaginary part of the dielectric function in Fig.[11-(a-ii), 13-(b-ii)] and the absorption coefficient in Fig.[10,12] are related to each other.

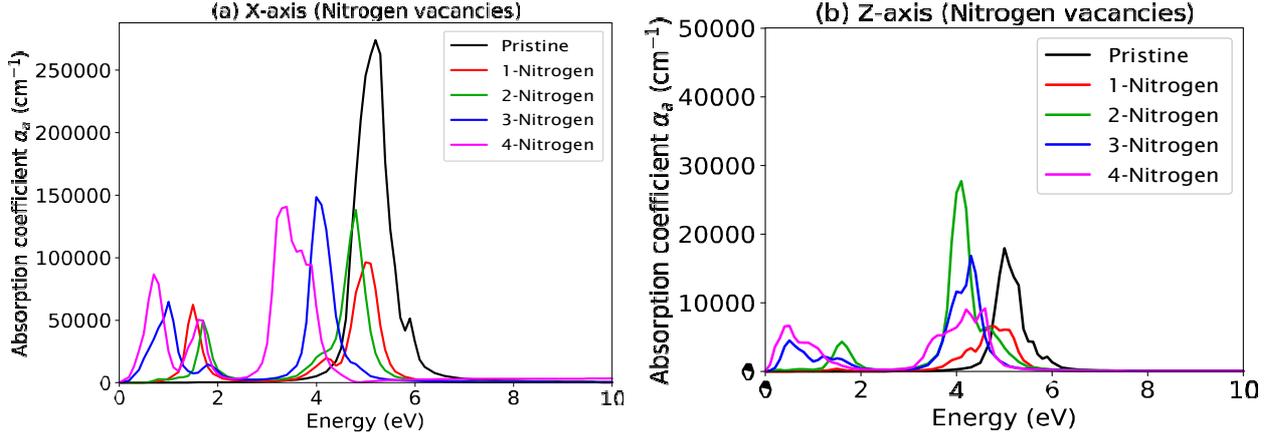

FIG. 12: Absorption coefficient for the Nitrogen vacant bilayer system with x=0.03125%-Fig(a), 0.0625%–Fig(b), 0.09375%-Fig(c) and 0.125%-Fig(d)vacancy is presented in the above figures.

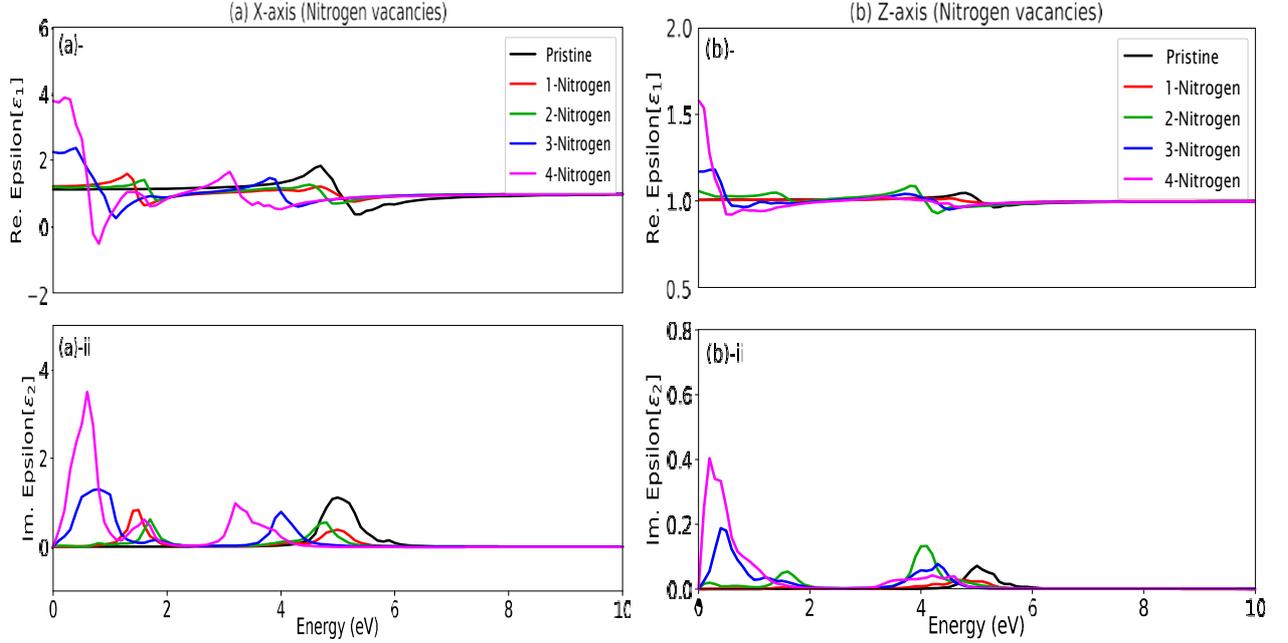

FIG. 13: Real and imaginary part of the calculated dielectric function of the Nitrogen vacant system are shown in the figures. Fig.(a) is for in plane symmetry i.e., along xx. Fig.(b) is for out of plane symmetry i.e., along zz.

The response of Nitrogen vacant system with light in terms of refractive index is presented in supporting figures [see supporting figures Fig-(a,b).16 along x and z axis labelled as $\eta_{xx}$ and $\eta_{zz}$.In case of 1-Nitrogen vacancy a peak is observed at 1.60eV. The value of $\eta_{xx}$ dips below 1 at 1.80eV and again at 4.8eV. In case of 2 Nitrogen Vacancies, the value of $\eta_{xx}$ dips down below once at 1.85eV and again at 4.65eV. Also for 3 and 4-Nitrogen vacant system $\eta_{xx}$ dips down twice below 1 at energies 1.0, 4.1eV and 0.65, 3.68eV respectively. The value of the refractive index for Nitrogen vacant system x=0.03125%, 0.0625%, 0.09375% and 0.125% becomes less than 1.00 at particular energy value which resembles transparent property of the material. The calculated vaue of refractive index$\eta_{zz}$ along zz axis for all B and N vacant systems is less in comparision with $\eta_{xx}$. The value of the refractive index $\eta_{xx}$ and $\eta_{zz}$ dips down below 1 at the same energy range for both B and N vacant systems [cf16].

## IV. CONCLUSION

We have calculated the electronic, magnetic and optical properties of bilayer BN with vacancy defects at B/N sites by using first principles method. The thermo-dynamic stability have been for the defective and pristine h-BN have been evaluated by calculating the phonon dispersion. The system are all dynamic stable with no imaginary modes. The electronic properties of pristine bilayer h-BN exhibit intrinsic semiconducting feature. The presence of defect vacancy in the B/N sites was observed to induce ferro-magnetism in the bilayers with a reduction in the electronic band gap. The net magnetic moment for the 1,2,3 and 4 bilayer h-Bn with boron and nitrogen vacancies obtained are 2.953, 3.144, 3.745 and 6.583$\mu_B$ and 0.903, 1.973, 2.975 and 3.926$\mu_B$ respectively, which arises due to the unpaired electrons of the unbonded Nitrogen and Boron atoms during h-BN vacancy at the B/N site. Our theoretical observation of the bilayer system with vacancies also gives a mechanism of tuning the band gap by creating B and N vacancy. The optical response of the pristine and defective systems show interesting properties which can be utilized in applications of optoelectronic devices. The refractive index and dielectric constant of the pristine bilayer h-BN and 1,2,3,4-B/N defective system show that the optical properties of the bilayer h-BN is highly senesitive and tunable. The 4-B and 4-N vacant system in the energy range 1.0-1.5eV and 0.677-0.897eV respectively can be utilised for making negative index optical materials. The results from the present work can be utilised in designing novel nano-electronic devices. Also, the co-existing of magnetic and semiconducting behaviour can have potential application in future data-transfer and data-storage technology.

## ACKNOWLEDGMENTS

D. P. Rai acknowledges Core Research Grant from Department of Science and Technology SERB (CRG DST-SERB, New Delhi India) via Sanction no.CRG/2018/000009(Ver-1).

———————————————

hexagonal boron Nitride. *APL Mater.* 7 021106 (2019). DOI:https://doi.org/10.1063/1.5087836

# V. SUPPORTING FILES

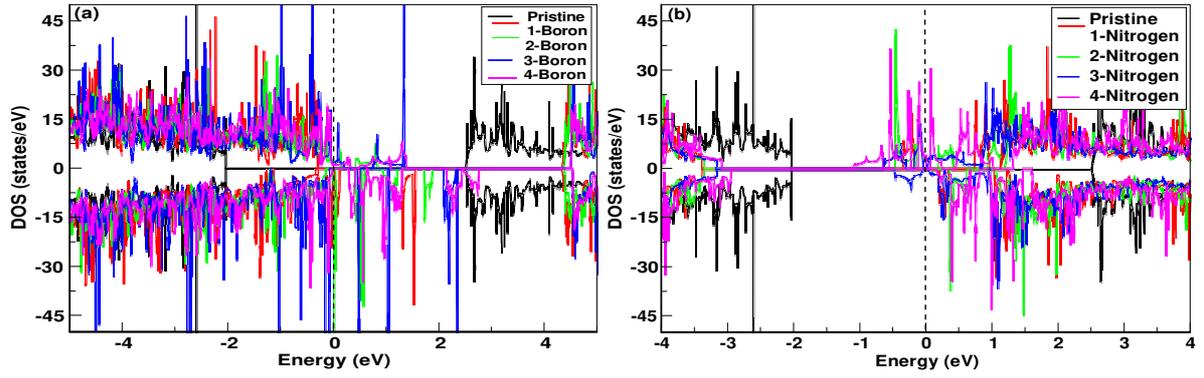

FIG. 14: Total spin-up and spin-down DOS of vacancy defects at: (a) B-site and (b) N-site with the Fermi level set to zero. Type of vacancy has been specified by the colour in both the figures.

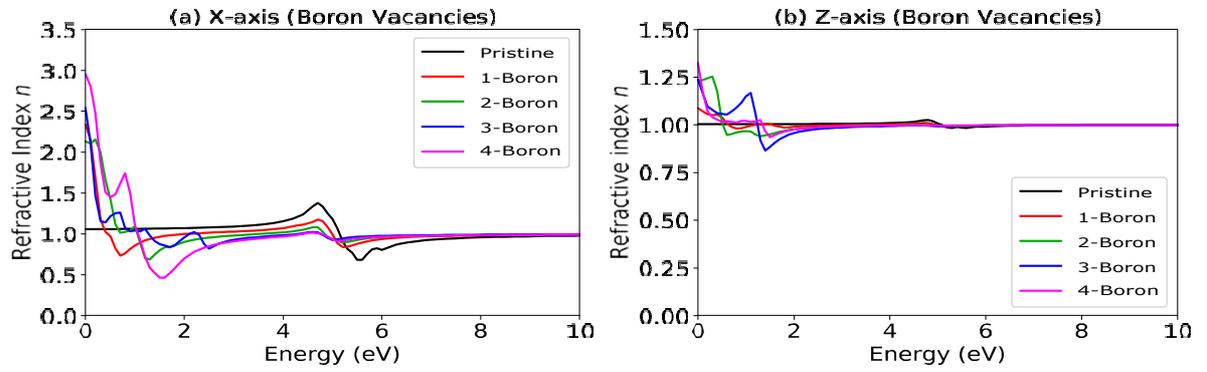

FIG. 15: In the above figure we have presented the calculated Refractive Index of Pristine and all the Boron vacant system. Fig(a)-Pristine,(b)-1-Boron(x=0.03125%) ,(c)-2-Boron(x=0.0625%),(d)-3-Boron(x=0.09375%),(e)-4-Boron(x=0.125%).

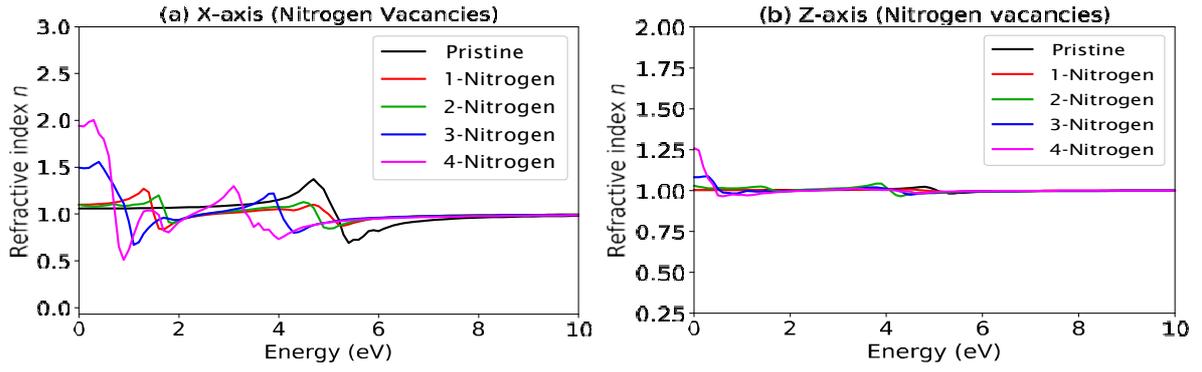

FIG. 16: In the above figure we have presented the calculated Refractive Index of all the Nitrogen vacant system. Fig(a)-1-Nitrogen(x=0.03125%),(b)-2-Nitrogen(x=0.0625%) ,(c)-3-Nitrogen(x=0.09375%),(d)-4-Nitrogen(x=0.125%) vacancy respectively.